\documentclass[preprint,showpacs,amsmath,amssymb,aip,cha]{revtex4-1}

\usepackage{graphicx}
\usepackage{dcolumn}
\usepackage{bm}
\usepackage{color}

\begin{document}

\title{Investigation of Near-Surface Defects of Nanodiamonds by High-Frequency EPR and DFT Calculation} 

\author{Z. Peng}
\affiliation{Department of Chemistry, University of Southern California, Los Angeles CA 90089, USA}

\author{T. Biktagirov}
\affiliation{Lehrstuhl f\"{u}r Theoretische Physik, Universit\"{a}t Paderborn,
Warburger Str. 100, 33098 Paderborn, Germany}

\author{F. H. Cho}
\affiliation{Department of Physics \& Astronomy, University of Southern California, Los Angeles CA 90089, USA}

\author{U. Gerstmann}
\affiliation{Lehrstuhl f\"{u}r Theoretische Physik, Universit\"{a}t Paderborn,
Warburger Str. 100, 33098 Paderborn, Germany}

\author{S. Takahashi}
\email{susumu.takahashi@usc.edu}
\affiliation{Department of Chemistry, University of Southern California, Los Angeles CA 90089, USA}
\affiliation{Department of Physics \& Astronomy, University of Southern California, Los Angeles CA 90089, USA}

\date{\today}

\begin{abstract}
Nanodiamond (ND) hosting nitrogen-vacancy (NV) centers is a promising platform for quantum sensing applications.
Sensitivity of the applications using NV centers in NDs is often limited due to presence of paramagnetic impurity contents near the ND surface.
Here, we investigate near-surface paramagnetic impurities in NDs.
Using high-frequency (HF) electron paramagnetic resonance spectroscopy, the near-surface paramagnetic impurity within the shell of NDs is probed and its $g$-value is determined to be 2.0028(3).
Furthermore, HF electron-electron double resonance-detected nuclear magnetic resonance spectroscopy and a first principle calculation show that a possible structure of the near-surface impurity is the negatively charged vacancy V$^-$.
The identification of the near-surface impurity by the present investigation provides a promising pathway to improve the NV properties in NDs and the NV-based sensing techniques.
\end{abstract}


\maketitle

\section{Introduction}
Diamond is a fascinating material, hosting nitrogen-vacancy (NV) defect centers with unique magnetic and optical properties.~\cite{Gruber97, Wrachtrup06}
In recent years, remarkable efforts have been put into studying fundamental quantum physics~\cite{Gaebel06, Epstein05, Childress06, Hanson08, Takahashi08, Wang13} and realizing applications to fundamental quantum information processing~\cite{Wrachtrup06, Dutt07, Fuchs09, Neumann10, Robledo11} as well as magnetic field sensing~\cite{Degen08, Balasubramanian08, Maze08, Taylor08, Maletinsky12, Grinolds13, Rondin14} using NV centers in diamond.
In NV-based magnetometry, spins inside diamond crystal ({\it e.g.}, $^{13}$C, single substitutional nitrogen defect centers, and other paramagnetic impurities)~\cite{Childress06, Dutt07, Neumann10, deLange12, Abeywardana14} as well as external spins in the vicinity of the surface of the diamond ({\it e.g.}, paramagnetic defects, radicals, $^1$H, Gd$^{3+}$, and Mn$^{2+}$)~\cite{Grinolds13, Grinolds14, Mamin12, Laraoui12, Tetienne13, Kaufmann13, Steinert13, Staudacher13, Mamin13sci, Sushkov14, Loretz14sci} have been successfully detected.
Difficulties in sensing external spins exist due to undesired spin and optical properties of NV centers ({\it e.g.}, short spin relaxation times and unstable photoluminescence) when NV centers are located close distance to diamond surface.~\cite{Tisler09, Tetienne13, Rosskopf14}
The origin of the undesirable properties is considered to be related to strain on NV centers and paramagnetic impurities existing near the surface.

There have been many reports that suggest the existence of specific paramagnetic impurities near surface of various kinds of diamonds.
Electron paramagnetic resonance (EPR) investigation of mechanically crushed diamonds revealed $g\sim2$ like signals that are attributed to structural damages near the diamond surface due to crushing process~\cite{Walters61} or $\sigma$-radicals.~\cite{Smirnov97}
EPR measurements of diamond powders produced by detonation process consistently have also shown $g\sim2$ like signals,~\cite{Iakoubovskii00, Shames02, Soltamova09, Dubois09} which are claimed to originate from dangling bonds associated with structural defects in the core or within the surface of diamond ({\it i.e.}, sp$^3$-hybridized carbon).
On the other hand, two separate nuclear magnetic resonance (NMR) studies of detonation diamond powders argue that paramagnetic impurities exist in a thin shell ($\sim$0.6 nm) near the surface,~\cite{Fang09} which is not associated with dangling bonds, or may be homogenously distributed throughout the whole volume of diamond crystal.~\cite{Dubois09}
Finally, studies of shallow NV centers in diamond crystals~\cite{Mamin12, Ofori-Okai12, Rosskopf14} as well as NV centers found in nanodiamond (ND) crystals~\cite{Tetienne13, Kaufmann13} have shown that these NV centers exhibit different spin properties ({\it e.g.}, broader linewidth and faster spin relaxation times) compared to deep, stable NV centers in diamond crystals, which are often explained by the existence of dense paramagnetic impurities on the surface of hosting diamonds.

In this article, we investigate near-surface defects and impurities in NDs.
We employ high-frequency (HF) (230 GHz and 115 GHz) and 9 GHz continuous-wave (cw) and pulsed EPR spectroscopy to study defect and impurity contents in various sizes of diamond crystals.
HF EPR spectroscopy is highly advantageous to distinguish paramagnetic centers existing in diamond with high spectral resolution.
230 GHz cw EPR spectra show the presence of two major impurity contents; single substitutional nitrogen impurity (P1 center) which is common in diamond, and paramagnetic impurity unique to NDs (denoted as X spin through this paper).
Moreover, particle-size dependence of the EPR intensity ratio between P1 and X spins indicates that X is localized in the vicinity of the diamond surface while P1 center is located in the core.
We also observe that the linewidth of X is much broader than that of P1 center, and further line broadening of X is visible as the electron Larmor frequency is increased from 9 GHz to 230 GHz.
We also study composition of X spin using hyperfine spectroscopy.
The technique we employ is electron-electron double resonance-detected nuclear magnetic resonance (EDNMR).
EDNMR is one of electron-electron double resonance techniques which excites two different electron spin transitions.~\cite{Schosseler94}
Compared with commonly used electron nuclear double resonance (ENDOR) spectroscopy which excites electron and nuclear spin transitions,
EDNMR has advantage in the signal sensitivity for a spin system with fast electron spin relaxations.
HF EDNMR also enables to achieve a high spectral resolution comparable to ENDOR.
With EDNMR investigation on the X spin where no signature of relevant hyperfine couplings are observed, we confirm that the X spin consists of neither hydrogen nor nitrogen atom.
Furthermore, we utilize a first principle calculation in the framework of density functional theory (DFT) to identify structures of the X spin.
The calculation result shows that a negatively charged vacancy-related defect is candidates of the X spin.

\section{MATERIALS AND METHODS}
\subsection{Diamond samples}
The investigation was performed with a single crystal (1.5$\times$1.5$\times$1.0 mm$^3$) type-Ib high-pressure high-temperature (HPHT) synthetic diamond (Sumitomo Electric Industries),
micron-size diamond powders ($10 \pm 1$ $\mu$m) (Engis Corporation), and eight different sizes of NDs (Engis Corporation and L. M. Van Moppes and Sons SA).
The mean diameters and standard deviations of NDs specified by the suppliers are $550 \pm 100$ nm, $250 \pm 80$ nm, $160 \pm 50$ nm, $100 \pm 30$ nm, $60 \pm 20$ nm, $50 \pm 20$ nm, and $30 \pm 10$ nm.
The 10-$\mu$m and ND powders were manufactured by mechanical milling or grinding of type-Ib diamond crystals
where the concentration of nitrogen related impurities in NDs is in the order of 10 to 100 parts per million (ppm) carbon atoms.

\subsection{HF EPR/EDNMR spectroscopy}
HF EPR and EDNMR experiments were performed using a home-built system at USC.
The system employs a high-power solid-state source consisting two microwave synthesizers (8-10 GHz and 9-11 GHz), pin switches, microwave amplifiers, and frequency multipliers.
For EDNMR measurement, a variable attenuator is implemented to control the power of the second HF microwave.
The output power of the HF source system is 100 mW at 230 GHz and 700 mW at 115 GHz.
The HF microwaves are propagated in free-space using a quasioptical bridge and then couple to a corrugated waveguide.
A sample placed on a metallic end-plate at the end of the waveguide, and then placed at the center of a 12.1 T cryogenic-free superconducting magnet.
In experiments on ND powders, ND powders ($\sim$5 mg typically) are placed in a teflon sample holder (5 mm diameter) and the teflon sample holder is placed on the end-plate.~\cite{Cho15}
EPR signals are isolated from the excitation using an induction mode operation.~\cite{Smith98}
For EPR/EDNMR experiment, we employ a superheterodyne detection system in which 115 GHz and 230 GHz is down-converted into an intermediate frequency (IF) of 3 GHz, and then again down-converted to in-phase and quadrature components of dc signals.
Both in-phase and quadrature signals are recorded to obtain the absorption and dispersion signals of EPR.
The microwave phase is adjusted to obtain correct shapes in both absorption and dispersion data.
Further details of the HF EPR/EDNMR system are described elsewhere.~\cite{Cho14, Cho15}
In the EPR/EDNMR measurements, the HF microwave power and the magnetic field modulation strength are adjusted carefully to maximize the intensity of EPR signals without distortion of the signals
(see Sect. S2 in supplemental materials for the power adjustment).
Typically modulation amplitude of 0.02 mT at modulation frequency of 20 kHz is used.

\subsection{X-band EPR spectroscopy}
X-band continuous-wave (cw) EPR spectroscopy was performed using an EMX system (Bruker Biospin).
For each measurement, samples were placed in a quartz capillary (inner diameter: 0.86 mm or 0.64 mm), with a typical sample volume being 1-5 $\mu$L.
cw EPR spectra are obtained with optimum microwave power and magnetic field modulation strength which maximize the amplitude of EPR signals without distorting the lineshape.
Typical parameter sets are a modulation amplitude of 0.03 mT and a modulation frequency of 100 kHz.

\section{RESULTS AND DISCUSSION }
\subsection{HF EPR spectroscopy: Detection and characterization of near-surface defects}
\begin{figure}
\includegraphics[width=90 mm]{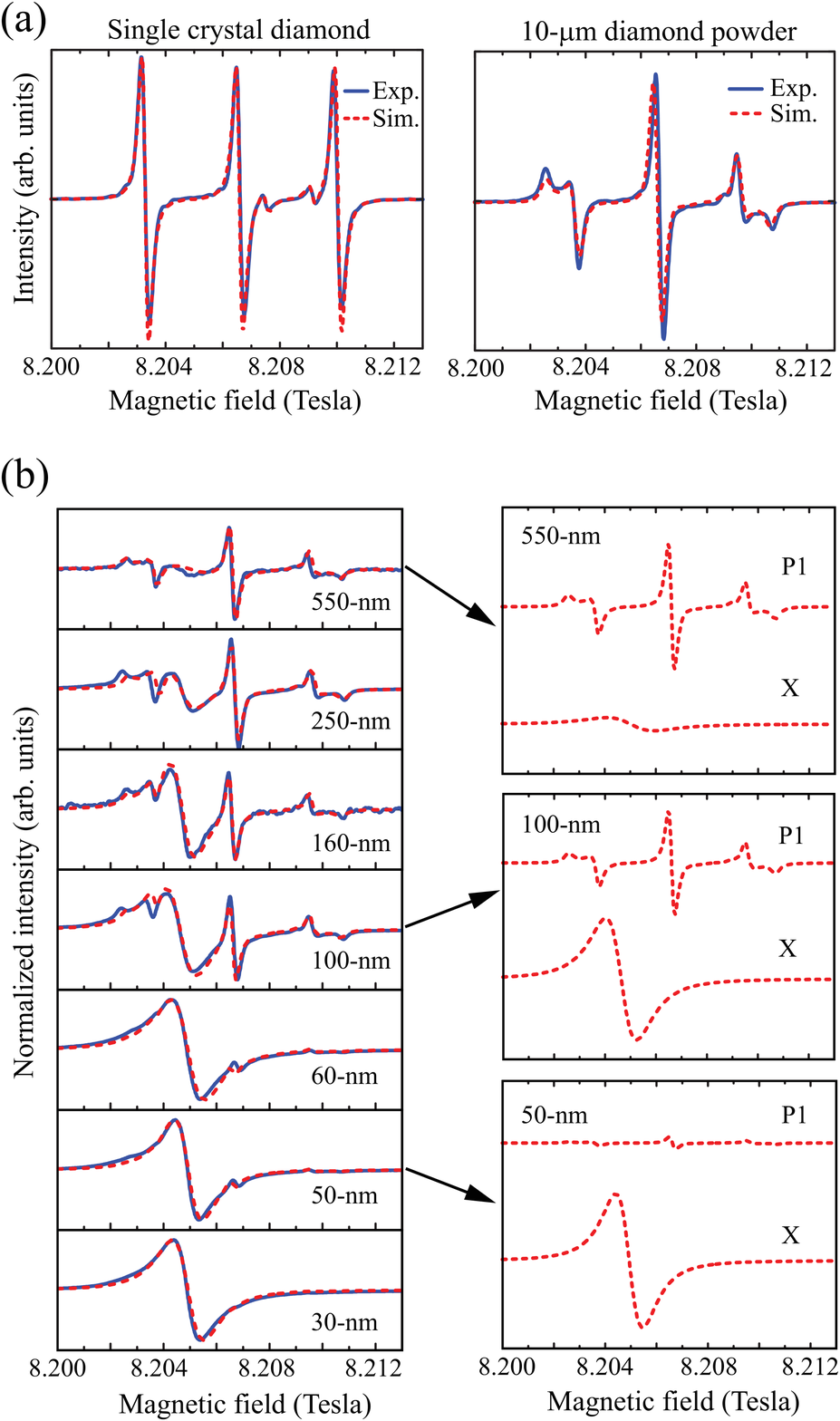}
\caption{
\label{fig:cwEPR}
230 GHz cw EPR spectra of the diamond samples.
(a) (Left) (100) single crystal diamond.
The external magnetic field ($B_{0}$) was applied along the $\langle 100 \rangle$ axis of the diamond.
(Right) Diamond powder with 10-$\mu$m mean diameter.
The signal is the so-called powder spectrum of P1 center.
(b) EPR spectra of all sizes of NDs.
(Left) The experimental and simulated spectra on NDs.
(Right) The experimental and simulated EPR spectra on 550-nm, 100-nm and 50-nm NDs.
The partial contributions from P1 and X spins are indicated by the red dashed lines.
All measurements were performed at room temperature.
The EPR spectrum analysis was done by Easyspin.~\cite{Easyspin}
}
\end{figure}
First, we discuss the study of paramagnetic impurity contents in the diamond samples using 230 GHz cw EPR spectroscopy.
Figure~\ref{fig:cwEPR}(a) shows 230 GHz EPR spectra of the single crystal diamond and 10-$\mu$m diamond powder samples taken using the HF EPR spectrometer.
As shown in Fig.~\ref{fig:cwEPR}(a), 230 GHz EPR spectrum of the single crystal diamond shows three pronounced signals from P1 centers.
The P1 center has $S=1/2$ and the hyperfine coupling to $^{14}N$ nuclear spin ($I=1$).
The spin Hamiltonian of P1 center is given by,
\begin{eqnarray}
\label{eq:Hn}
H_{N}=\mu_B g_{N} {\bm B_0} \cdot {\bm S^{N}} + {\bm S^{N}} \cdot \stackrel{\leftrightarrow}{A} \cdot {\bm I^{N}} + P_z(I_z^{N})^2,
\end{eqnarray}
where $\mu_{B}$ is the Bohr magneton, $g_{N}=2.0024$ is the isotropic $g$-value of P1 center, ${\bm B_{0}}$ is the external magnetic field, ${\bm S^{N}}$ and ${\bm I^{N}}$ are the electron and nuclear spin operators, respectively.
$\stackrel{\leftrightarrow}{A}$ is the anisotropic hyperfine coupling to $^{14}N$ nuclear spin ($A_{x,y}=82$ MHz and $A_z=114$ MHz).~\cite{Loubser78, Takahashi08}
The nuclear quadrupole coupling $P_z=-4$ MHz.~\cite{Cook66}
As shown in Fig.~\ref{fig:cwEPR}(a), EPR spectrum of the single crystal diamond was simulated using the P1 spin Hamiltonian (Eq.~(\ref{eq:Hn}))
and we found a good agreement between the observed EPR signal and the simulated spectrum.
In addition, EPR spectrum of the 10-$\mu$m diamond powder is shown in Fig.~\ref{fig:cwEPR}(a).
The powder sample contains ensembles of diamond crystals which are randomly oriented with respect to ${\bm B_{0}}$,
therefore all the orientations of P1 centers were taken into account to obtain so-called powder spectrum.
As shown in Fig.~\ref{fig:cwEPR}(a), the simulated powder spectrum also agrees well with the observed EPR signal.

Next, we discuss the size dependence of EPR spectra on the ND samples.
Figure~\ref{fig:cwEPR}(b) shows 230 GHz EPR spectra of NDs with mean diameters from 550 nm to 30 nm.
As shown in Fig.~\ref{fig:cwEPR}(b), 230 GHz EPR spectroscopy enabled to resolve two EPR signals in the ND samples;
(i) one is EPR signal of P1 centers which was also observed in the single crystal diamond and 10-$\mu$m powder samples.
(ii) the other is the EPR signal at 8.2047 Tesla (denoted as X in Fig.~\ref{fig:cwEPR}(b)).
As shown in Fig.~\ref{fig:cwEPR}(b), the EPR intensities of P1 and X spins largely depend on the size of NDs, {\it i.e.} for P1 centers, larger the size of NDs is, stronger the EPR intensity is, and, for X spins, smaller the size of NDs is, stronger the EPR intensity is.
We also noticed that the X contribution is well represented by a single $S=1/2$ EPR signal.
Therefore, in order to simulate the observed EPR spectra of X spins, we considered the spin Hamiltonian for $S=1/2$ with $g_X=2.0028$.
By considering EPR spectra for P1 (Eq.~(\ref{eq:Hn})) and X spins,
we found that the observed EPR data can be explained very well for all investigated ND sizes.

\begin{figure}
\includegraphics[width=160 mm]{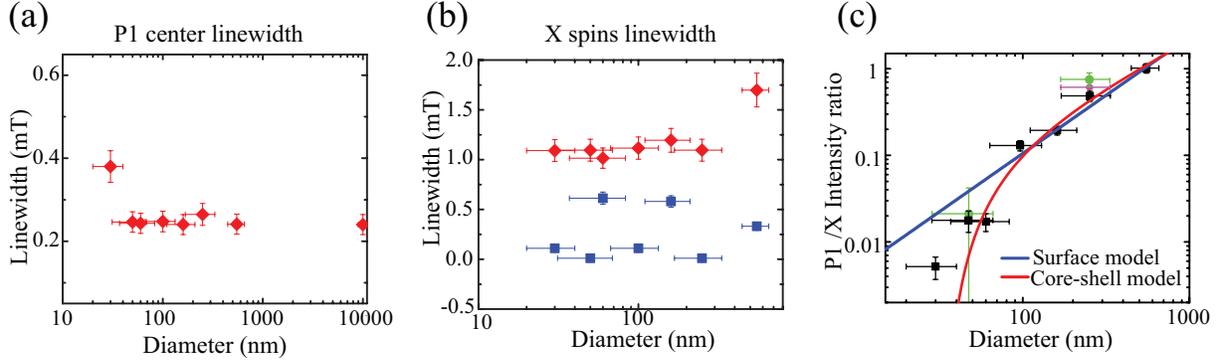}
\caption{
\label{fig:cwEPRsum}
Size dependence of EPR linewidth and intensity of P1 centers and X spins.
(a) Linewidth of P1 centers as a function of the diamond size.
Lorentzian linewidth (red diamond) was obtained from the fit.
(b) Linewidth of X spins as a function of the diamond size.
The Voigt function was used in the fit and the peak-to-peak Lorentzian (red diamond) and Gaussian (blue square) linewidths were obtained.
(c) Intensity ratio of P1 centers and X spins as a function of the diamond size.
The relative intensity ratio of P1 centers and X spins from 230 GHz EPR is shown by the black solid squares.
The green and purple solid circles represents the intensity ratio data obtained from X-band and 115 GHz spectra, respectively.
The blue solid curve shows the fit result using a simple surface model (P1$/$X intensity $\sim$ diameter) while the red solid curve shows the fit result using the core-shell model.
Overall, the core-shell model gives a better fit.
}
\end{figure}
We also analyzed the EPR intensity of P1 and X spins.
The intensity ratios of P1 and X spins were obtained from the fit of the experimental spectrum to calculated EPR spectra of P1 and X spins.
In the fit, the intensity and linewidths were fitting-parameters, and their errors (95 $\%$ confidence interval) were also obtained from the fit.
Figure~\ref{fig:cwEPRsum} shows the result of cw EPR analysis.
The analysis shows that the P1 lineshape is dominated by the Lorentzian contribution.
As shown in Fig.~\ref{fig:cwEPRsum}(a),  the peak-to-peak linewidth of the Lorentzian lineshape is independent of the size of NDs.
As shown in Fig.~\ref{fig:cwEPRsum}(b), the lineshape of the X spins is well explained by the Voigt function.
From the analysis, we found that the ratio of the contributions is independent of the ND size and their peak-to-peak linewidths in Lorentzian and Gaussian components are still independent of the size of NDs.
Furthermore, from the result of the lineshape analysis, we extracted the cw EPR intensity ratio between P1 and X spins.
Figure~\ref{fig:cwEPRsum}(c) shows the EPR intensity ratio ($I_{P1}/I_{X}$) as a function of the size of NDs.
Observation of strong size dependence on EPR intensity ratio indicates that X is localized in the vicinity of the surface of ND crystals.
In order to explain the size dependence, we consider the core-shell model.
In the core-shell model, X spins are located in the spherical shell of thickness $t$ from the near-surface ({\it i.e.} shell) region while P1 centers are only located in the core of NDs, therefore, $I_{P1}/I_{X} \sim [(4/3 \pi (r-t)^3)]/[(4/3 \pi r^3 - 4/3 \pi (r-t)^3)]$ where $r$ is the radius of NDs.
The spin concentration ratio between P1 and X spins is assumed to be same for different sizes of NDs.
As shown in Fig.~\ref{fig:cwEPRsum}(c), we found good agreement of the size dependence data with the core-shell model.
From the fit, we also obtained an estimate of $9\pm2$ nm for the shell thickness $t$.

\begin{figure}
\includegraphics[width=150 mm]{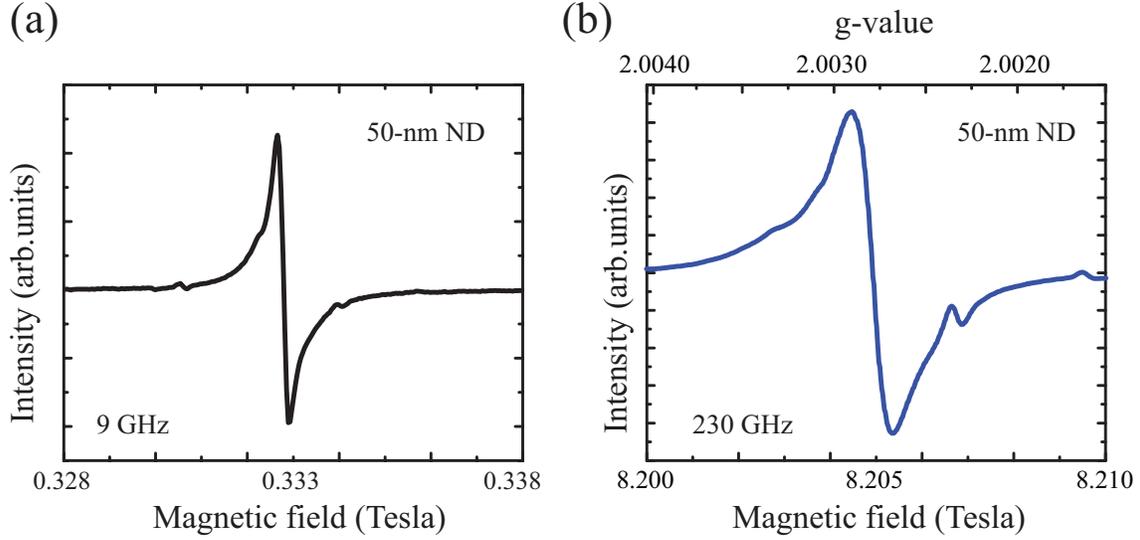}
\caption{
\label{fig:FreqDep}
EPR spectra on 50-nm ND taken by the X-band and 230 GHz EPR spectrometers.
(a) EPR spectrum taken at 9 GHz (X-band EPR). The peak-to-peak linewidth is 0.24 mT.
(b) EPR spectrum taken at 230 GHz. The peak-to-peak linewidth is 1.00 mT.
The x-ranges of both data are 0.01 Tesla.
}
\end{figure}
Furthermore, we investigated frequency dependence (X-band, 115 GHz and 230 GHz) of EPR spectra with 50-nm and 250-nm NDs.
Figure~\ref{fig:FreqDep} shows EPR spectra of the 50-nm ND sample taken at 9.3 GHz and 230 GHz where the EPR signal of the 50-nm ND sample is dominated by X spins.
As shown in Fig.~\ref{fig:FreqDep}, the EPR linewidths at 9.3 GHz and 230 GHz are clearly different.
The observation indicates that the origin of the broadening is related to $g_X$-value, \textit{i.e.}, $g$-strains.
By considering the full-width at half-maximum of the EPR spectrum (Fig.~\ref{fig:FreqDep}(b)), we estimated the distribution of the $g$-value ($\Delta g_X$) to be $\pm 0.0003$, {\it i.e.} $g_X = 2.0028 \pm 0.0003$.
The employment of HF EPR was critical for the identification of X-spins in this experiment
because of the small difference in their g-values which causes a significant overlap in X-band spectrum (Sect. S3 in supplemental materials).
The EPR intensity ratio between P1 and X spins was also analyzed using spectra from 50-nm and 250-nm NDs (Fig.~\ref{fig:FreqDep} and Fig. S3).
As shown in Fig.~\ref{fig:cwEPRsum}(c), the result of the size dependence does not depends on the EPR frequency, however, the errors in the 230 GHz EPR analysis are significantly smaller because of the spectral distinction of P1 and X spins at 230 GHz EPR.

\subsection{HF EDNMR spectroscopy: Investigation of near-surface impurity structures}
\begin{figure}
\includegraphics[width=140 mm]{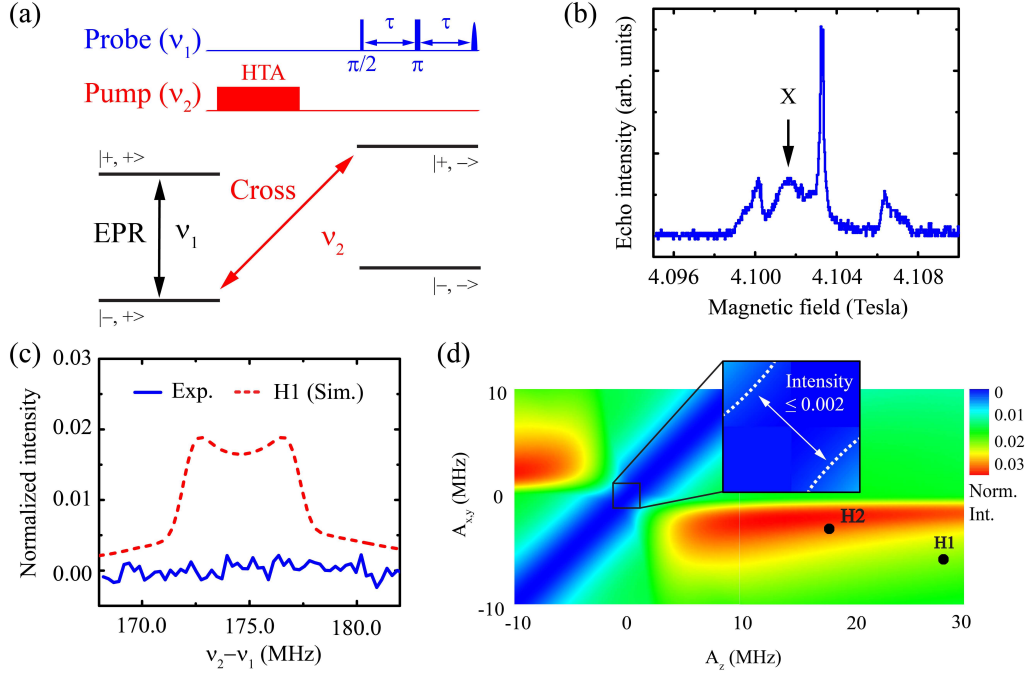}
\caption{
\label{fig:EDNMR_H}
Proton EDNMR experiment using X spin. (a) Overview of EDNMR experiment.
EDNMR pulse sequence consists of pulses with two microwave frequencies.
A high turning angle (HTA) pump pulse at the frequency of $\nu_{2}$ induces the population inversion of the cross transition.
A change of the population is detected via the spin echo sequence at the frequency of $\nu_{1}$.
(b) Echo detected EPR on 50-nm NDs taken at room temperature.
The black arrow points the echo signal from X spins.
The pulse parameters are $\pi /2 = 150$ ns, $\pi = 200$ ns, $\tau = 320$ ns and repetition time$~= 10$ ms.
(c) EDNMR experimental data of X spins (blue solid line) and the simulated EDNMR spectrum of H1 defects (red dashed line).
The x axis is frequency offset ($\nu_{2} - \nu_{1}$ and the y axis is EDNMR intensity normalized by the echo intensity without the HTA pulse.
Experimental parameters are $\pi /2 = 150$ ns, $\pi = 200$ ns, $\tau = 320$ ns, HTA pulse amplitude $w_{1} = 3.33$ MHz, HTA$~= 100$ $\mu$s and repetition time$~=10$ ms.
The simulation parameters are $A_{x,y} = -5.5$ MHz, $A_{z} = 27.5$ MHz, $w_{1} = 3.33$ MHz, $T_{2} = 153$ ns and HTA$~= 100$ $\mu$s.
(d) The simulated EDNMR peak intensity as a function of the hyperfine coupling constants.
The ranges of $A_z$ and $A_{x,y}$ are $-10$ to 30 MHz and $-10$ to 10 MHz, respectively.
The hyperfine couplings of H1 and H2 defects are indicated by black dots.
The inset shows a zoom-in image of (d) where the range of $A_z$ and $A_{x,y}$ are from - 1 to 1 MHz.
The two white dashed lines indicates the hyperfine couplings corresponding to the observed noise level ($0.2$ $\%$) in the experiment.
}
\end{figure}
Next, we discuss HF EDNMR experiment.
Identification of the composition of X spin is imperative.
Previous studies indicated that there exist dangling bonds and hydrogen and nitrogen-related defects near the diamond surface.~\cite{Zhou96, Loubser78}
Therefore, the aim of the EDNMR experiment is to detect hyperfine couplings from proton and nitrogen nuclear spins.
Fundamentals of the EDNMR measurement is described in Fig.~\ref{fig:EDNMR_H}(a) using a $S = 1/2$ electron spin system coupled to an $I = 1/2$ nuclear spin with a weak hyperfine interaction ($\omega_{NMR} > A_z$).
As shown in the pulse sequence of the EDNMR (Fig.~\ref{fig:EDNMR_H}(a)), the experiment is started with a high turning angle (HTA) pulse to excite the cross transition, then EDNMR signal is detected by a change of the echo intensity due to the population inversion induced by the HTA pulse.
Since the resonant frequency of the cross transition and effectiveness of the population inversion by the HTA pulse depend on the hyperfine coupling strength, the detection of EDNMR spectrum allows us to probe and measure the hyperfine coupling strengths.

In the experiment on NDs, we first performed an echo-detected field sweep measurement at 115 GHz to determine the resonance field of the X spin.
As shown in Fig.~\ref{fig:EDNMR_H}(b), the data clearly shows the signal from X spins at 4.1017 Tesla.
Then, we performed the EDNMR experiment at 4.1017 Tesla (with $\nu_1 = 115$ GHz).
Figure~\ref{fig:EDNMR_H}(c) is the experimental result which shows no visible EDNMR signal.
The noise level of the measurement was estimated to be $\sim 0.2 \%$.
The previous study on the diamond surface defects~\cite{Zhou96} reported two hydrogen-related defects called H1 ($S=1/2$, $g=2.0028$, $I=1/2$ and $A_{x,y} = -5.5$ MHz and $A_{z} = 27.5$ MHz) defects and H2 ($S=1/2$, $g=2.0028$, $I=1/2$ and $A_{x,y} = -2.7$ MHz, $A_{z} = 17.4$ MHz) defects.
H1 defect was also observed by other studies.~\cite{Watanabe88, Fanciulli93, Jia93, Holder94}
To compare with the experimental result with an expected EDNMR spectrum of H1, we perform simulation of EDNMR signals using Easyspin (the simulation procedure is described elsewhere~\cite{Cox17}).
As shown in Fig.~\ref{fig:EDNMR_H}(c), the simulated spectrum for H1 defects has much higher intensity than the observed noise.
In addition, the simulated spectrum for H2 defects has even higher EDNMR intensity.
Therefore, our analysis strongly suggests that X spin is not hydrogen-related defect (see Sect. S5 in supplemental materials for additional supporting information).
Furthermore, a contour plot in Fig.~\ref{fig:EDNMR_H}(d) shows the simulated EDNMR peak intensity as a function of hyperfine coupling strengths where $A_z$ and $A_{x,y}$ are considered from $-10$ to 30 MHz and from $-10$ to 10 MHz, respectively.
As shown in Fig.~\ref{fig:EDNMR_H}(d), when the hyperfine coupling is zero or isotropic, the EDNMR intensity also becomes zero.
On the other hand, the intensity of an anisotropic hyperfine coupling increases, EDNMR intensity also increases.
Based on the observed noise level, we estimated detectable hyperfine couplings in Fig.~\ref{fig:EDNMR_H}(d) (the white dashed line in the figure) with which the EDNMR intensity becomes the noise level.

\begin{figure}
\includegraphics[width=80 mm]{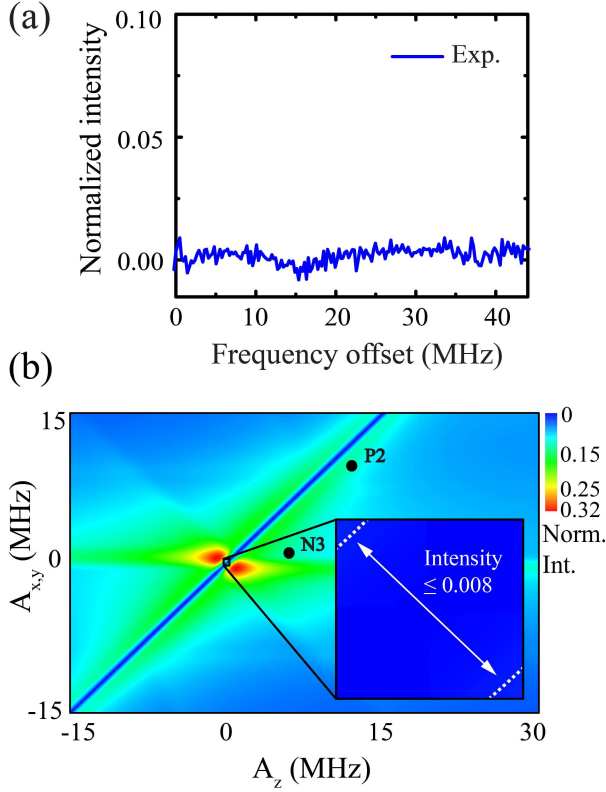}
\caption{
\label{fig:EDNMR_N}
$^{14}N$ EDNMR.
(a) EDNMR experimental data.
The noise level was estimated to be $\sim$ 0.8 $\%$.
Experimental parameters are $\pi /2=150$ ns, $\pi = 200$ ns, $\tau = 320$ ns, $w_{1} = 3.33$ MHz, HTA$~= 100$ $\mu$s and repetition time$~=10$ ms.
(b) Simulated EDNMR intensity as a function of the hyperfine couplings.
The hyperfine constants of P2 and N3 were indicted in the figure.
The inset shows the simulated EDNMR for hyperfine coupling below 0.05 MHz.
EDNMR signal corresponding to the intensity of $<$ 0.8 $\%$ (the noise level) is indicated by the white dashed line.
}
\end{figure}
Next, we discuss EDNMR experiment to detect a $^{14}N$ hyperfine coupling.
There exist many nitrogen-related impurities in diamond.~\cite{Loubser78}
Among those impurities, we consider the following $S=1/2$ systems because of their $g$-values and hyperfine couplings consistent with EPR spectrum of X spin.
(1) P2 (consisting of three nitrogen atoms with $g=2.003\pm0.001$, $I=1 (^{14}N)$, $A_{x,y}=10.10$ MHz, $A_{z}=11.00$ MHz for all nitrogen nuclear spins);
(2) N3: (consisting of two vacancies and one nitrogen atom with $g=2.003$, $I=1  (^{14}N)$, $A_{x,y}=1.50$ MHz, $A_{z}=5.10$ MHz).
In order to detect the hyperfine couplings of $^{14}N$, we performed EDNMR experiment in the frequency range of $^{14}N$ NMR.
As shown in Fig.~\ref{fig:EDNMR_N}(a), the experimental result shows noise level $\sim 0.8$ $\%$ and no visible NMR signal from $^{14}N$.
Based on the simulated EDNMR spectra with the hyperfine couplings listed above, those two nitrogen centers are expected to give much higher EDNMR intensities than the noise level as indicated in Fig.~\ref{fig:EDNMR_N}(b).
Therefore, the EDNMR result excludes nitrogen-related impurities for X spins.
Based on the detected noise level of the experiment, detectable hyperfine couplings in the present EDNMR experiment are indicated in Fig.~\ref{fig:EDNMR_N}.
Overall, the HF EDNMR experimental results suggest that the X spin is a vacancy-related defect.

\subsection{DFT calculation: Identification of near-surface impurities}
Finally, we discuss possible structures of the near-surface vacancy-related defect.
For the investigation, we employ a first principles calculation in the framework of density functional theory (DFT) to identify paramagnetic impurities consistent with the observed EPR spectrum.
A direct {\em ab initio} treatment of the entire volume of a nanoparticle, for example, with a diameter of 30 nm, requires a DFT modeling for several ten thousands of atoms.
Despite the ongoing progress on high performance computing (HPC), the corresponding computational costs for the ND calculations still exceed by far nowadays available HPC resources.
In this work, we therefore focus the investigation on the vicinity of the ND surface (the shell of ND) by considering a small volume with up to 250 atoms.
In the calculation, an irregularly formed ('{\em potato}'-like) volume containing 200 C atoms is initially cut from the diamond crystal.
We next perform molecular dynamics (MD) calculations under admixture of the diamond lattice and hydrogen atoms to find an optimum shape and surface from the DFT model.
As a result, we found that dangling bonds at the diamond surface tend to be passivated by dimerization of carbon atoms (surface reconstruction) and by hydrogen termination.
Additionally, when a single carbon atom exists on the diamond surface, the carbon atom is removed from the diamond surface with formation of CH$_4$ molecule.
\begin{figure}
\includegraphics[width=150 mm]{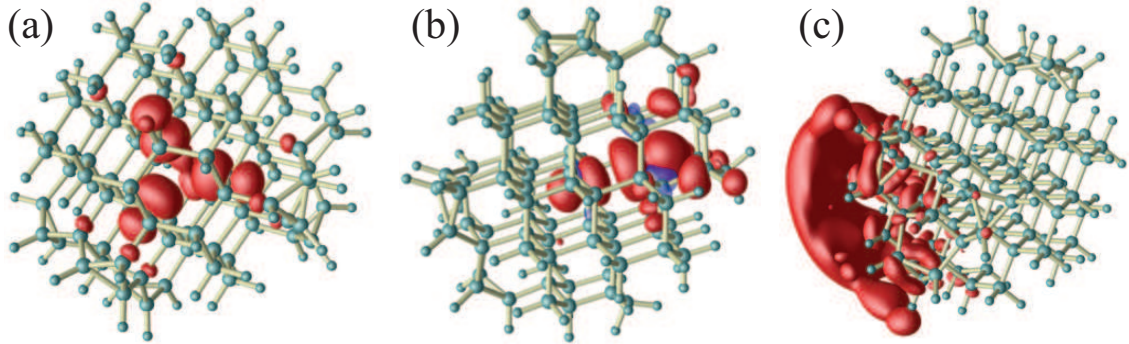}
\caption{
\label{fig:DFT}
Isosurfaces of the calculated spin densities.
The EPR properties arises from the magnetization density $m(r) = n^\uparrow(r)-n^\downarrow(r)$.
(a) The negative vacancy V$^-$ in a minimum-size ND.
(b) Substitutional N$_{\rm C}$ (the usual P1).
(c) Substitutional N$_{\rm C}$ in the more stable N$_{\rm C}^+$+e$^-$ configuration where the unpaired electron is transferred to the surface.
}
\end{figure}
After this MD treatment, the surface of the resulting NDs are found to be completely terminated by H atoms.
Figure~\ref{fig:DFT} shows a shell-only ND containing in total 260 atoms, 190 carbon and 70 hydrogen atoms.
Single vacancy and nitrogen-related defects (by taking out selected C atoms and/or substituting them by N atoms) have been already intentionally introduced as shown in Fig.~\ref{fig:DFT}.
In this way, the created structures are fully relaxed in a few different charge states.
We then calculate EPR parameters for the resulting spin-systems using the GIPAW pseudopotential formalism~\cite{Pickard02, Biktagirov18} implemented in the Quantum ESPRESSO package~\cite{Giannozzi09,Giannozzi17} (see supplemental material for computational details and comparative data for single crystal diamond).
The resulting DFT-calculated $g$ tensors for the most convenient structures in ND are compiled in Table I.

\begin{table}[ttt]
\caption{
\label{tab:g_tensors}
Calculated $g$ tensors for vacancy-related defects, isolated N-impurities (P1 centers) and NV-centers in 'shell-only' NDs ({\it cf.} Fig.~\ref{fig:cwEPRsum}) compared with the experimental value for the X spins.
}
\begin{tabular}{l c c c c c}
\hline\hline
system & \hspace*{0.5cm}$g_x$\hspace*{0.5cm}
       & \hspace*{0.5cm}$g_y$\hspace*{0.5cm}
       & \hspace*{0.5cm}$g_z$\hspace*{0.5cm}
       & $\bar{g}$
       & \hspace*{0.5cm}$D$ (MHz)\hspace*{0.5cm} \\ \hline
X-center (Exp.)          &          &          &         & 2.0028 & \\
V$^{2-}$ ($S$=1)         & 2.00284  & 2.00301  & 2.00310 & 2.00299 & -81 \\
V$^{-}$  ($S$=3/2)       & 2.00267  & 2.00278  & 2.00283 & 2.00276 & -32 \\
V$^{0}$  ($S$=1)         & 2.00228  & 2.00241  & 2.00314 & 2.00261 & 6053 \\
\hline\\
P1 center (Exp.)       &          &          &         & 2.0024  & \\
P1     ($S$=1/2)       & 2.00230  & 2.00241  & 2.00244 & 2.00238 & --- \\
N$^+$+e$^-$ ($S$=1/2)  & 2.00232  & 2.00232  & 2.00232 & 2.00232 & --- \\
NV$^-$ ($S$=1)         & 2.00263  & 2.00266  & 2.00297 & 2.00275 & 2830 \\
NV$^0$ ($S$=1/2)       & 2.00223  & 2.00257  & 2.00388 & 2.00289 & --- \\
\hline \hline
\end{tabular}
\end{table}

Among calculated vacancy-related defects, Table I contains indeed some defects with the $g$-values consistent with the experiment ($g = 2.0028 \pm 0.0003$).
In particular, the negatively charged vacancies V$^-$ and V$^{2-}$ provide $g$ tensors in good agreement with the $g$-value obtained from the experiment.
The averaged $g$-values and the anisotropy of V$^-$ and V$^{2-}$ are close to the experimental value.
In both single crystal material and modeled ND, the V$^-$ defect gives rise to a $S=3/2$ high-spin ground state (${\bar g}=2.00276$).
While the zero-field splitting (ZFS) of V$^-$ is exactly zero from symmetry reasons in a case of single crystal diamond,
in the shell region of NDs, the symmetry is lifted by local strain and anisotropic distortions.
A calculated value of ZFS for V$^-$ is less than $D=-30$ MHz.
The obtained $g$- and $D$-values for V$^-$ are consistent with the experimentally observed EPR position and linewidth.
In contrast, for the twofold negatively charged vacancy V$^{2-}$, the $g$-value of 2.00299 appears slightly too high.
In addition, the symmetry reduction within the ND reduces the D-value from $-143$ MHz in single crystal material
(V$^{2-}$ in D$_{2d}$ symmetry), but the resulting D values of at least $-80$ MHz are still too large to be covered by the observed EPR linewidth of the X spins.
In addition, although the defect with the neutral charge state (V$^0$) has a $g$-value comparable with the experiment ($\bar g =2.00261$), the calculated zero-field splitting shows $D=6.05$ GHz which should be clearly visible in the experiment.
Therefore, V$^0$ is no structure of the observed X spin.
Additionally, divacancies and trivacancies with various charge states were also considered in the DFT calculation.
However, we found that the resulting structures have ${\bar g}$ values below 2.0025 and too large $g$-tensor anisotropies for X spin as well.
Therefore, divacancies and trivacancies are also not the structures.

Furthermore, we performed DFT calculation on nitrogen-related defects in NDs.
Usually, the unpaired electron of substitutional N atoms in diamond tends to remain near the defect.
For example, the electronic and magnetic properties of a substitutional nitrogen defect P1 center are predominantly determined by its $p$-like unpaired electron, leading to an off-centered position of the nitrogen atom whereby the bond length to one of the four carbon ligands is increased by about 30$\%$.~\cite{Tucker94}
The present DFT method enables to calculate this configuration.
The DFT calculated $g$-value of 2.00238 is in very good agreement with the experimentally observed value for the P1 centers in the core region (see Table I).
On the other hand, when the N atom is located close to the surface, its unpaired electron tends to be released.
It can be transferred to the surface and distributed within an electron cloud located 2 to 4 {\AA} above the surface terminating atoms (see Fig.~\ref{fig:DFT}(c)), thereby showing free-electron like behavior (isotropic $g$-tensor with $g_e=2.002319$).
In comparison to the calculated P1-like configuration, about 0.3 eV are gained in the substitutional N defect.
Alternatively, the unpaired electron can be trapped by other defects, {\it e.g.} vacancies resulting in negatively charged V$^-$ and V$^{2-}$ discussed above.
In those cases, the substitutional nitrogen defect itself is effectively incorporated in ionized N$^+$ form and is not EPR-active anymore.
This is consistent with the experimental observation where P1 EPR signal is significantly suppressed in small NDs (see Fig. 2).
In parallel, the scenario of electron transfer from ionized P1 to vacancies supports near-surface negatively charged V$^-$ (V$^{2-}$) as structures responsible for the X spins.

Furthermore, we briefly note that NV-type defects have to be ruled out from structures of X spins.
For $S=1$ NV center (NV$^-$) in ND, the calculated zero-field splitting ($D=2.83$ GHz) is much larger than the observed EPR spectrum shown in Fig.~\ref{fig:cwEPR}.
For the neutral NV$^0$ ($S=1/2$) in ND, the calculated $g_z$ component (2.00388) is inconsistent.
Furthermore, the calculated $^{14}$N hyperfine constant in NV$^0$ is $\approx$ 9 MHz, which is two orders of magnitude large than the estimated detection limit of the present EDNMR experiment and such the hyperfine coupling should be visible ({\it cf.} Fig.~\ref{fig:EDNMR_N}(b)).
Therefore, both NV$^0$ and NV$^-$ have to be ruled out from structures of X spin.

\section{Summary}
In summary, we investigated near-surface paramagnetic defects in NDs using HF EPR and EDNMR spectroscopy, and DFT calculation.
The HF EPR studies probed near-surface paramagnetic defects in NDs.
The $g$-value of the near-surface defects was determined to be 2.0028(3).
With the assumption of the spin concentration ratio between P1 and X spins to be independent of the ND size,
the localization of X spins can be well explained by the core-shell model with the shell thickness of $9\pm2$ nm.
HF EDNMR spectroscopy was employed to investigate the physical structures of X spins where no hyperfine coupling with hydrogen and nitrogen nuclear spins was observed.
Those results confirmed that X spins are not dominated by hydrogen and nitrogen-related impurities and most likely they are vacancy-related defects.
Furthermore, the DFT study showed that the most probable structure behind the X spins is the negatively charged monovacancies V$^-$.
Based on the fabrication in which NDs are created by milling of type-Ib crystalline diamond crystals and no NMR signals obtained from EDNMR,
we speculate that X-spins are related to lattice defects which are specific to NDs fabricated by the milling process.
Quantum coherence of NV centers, which is important for NV-based sensing techniques, is often limited by surrounding paramagnetic defects and impurities.
The identification of the near-surface paramagnetic defects by the present investigation provides an important clue for improvement of the NV properties in NDs.

\section{Supplemental materials}
See supplemental materials for details of the HF EPR/EDNMR system and experiment, the AFM data and the DFT calculation.

\section{Acknowledgement}
This work was supported by the National Science Foundation (DMR-1508661 and CHE-1611134), the USC Anton B. Burg Foundation and the Searle scholars program (ST), and Deutsche Forschungsgesellschaft (DFG, via priority program SPP-1601) (UG). The numerical calculations have been done at the Paderborn Center for Parallel Computing (PC$^2$).

%

\end{document}